\documentclass[journal]{IEEEtran}
\ifCLASSINFOpdf
   \usepackage[pdftex]{graphicx}
\else
\fi
\usepackage{amsmath}
\usepackage{hyperref}
\usepackage{url}

\usepackage{cite}

\usepackage{multirow}
\usepackage{multicol}
\usepackage{mathtools}

\usepackage{amsmath,epsfig}
\usepackage{amsfonts}
\usepackage{url}
\usepackage{amssymb}
\usepackage{amsthm}
\newtheorem{definition}{Definition}[section]

\usepackage{caption}
\usepackage{subcaption}

\begin{document}
%
\title{Permutation invariance and uncertainty in multitemporal image super-resolution}
%
%
%

\author{Diego~Valsesia, \IEEEmembership{Member, IEEE}
        and~Enrico~Magli, \IEEEmembership{Fellow, IEEE}%
        \thanks{The authors are with Politecnico di Torino -- Department of Electronics and Telecommunications, Italy. email: \{name.surname\}@polito.it.}
}

%
%

\markboth{}%
{Valsesia \MakeLowercase{\textit{et al.}}: Permutation invariance and uncertainty in multitemporal image super-resolution}
%



\maketitle

\begin{abstract}
Recent advances have shown how deep neural networks can be extremely effective at super-resolving remote sensing imagery, starting from a multitemporal collection of low-resolution images. However, existing models have neglected the issue of temporal permutation, whereby the temporal ordering of the input images does not carry any relevant information for the super-resolution task and causes such models to be inefficient with the, often scarce, ground truth data that available for training. Thus, models ought not to learn feature extractors that rely on temporal ordering. In this paper, we show how building a model that is fully invariant to temporal permutation significantly improves performance and data efficiency. Moreover, we study how to quantify the uncertainty of the super-resolved image so that the final user is informed on the local quality of the product. We show how uncertainty correlates with temporal variation in the series, and how quantifying it further improves model performance. Experiments on the Proba-V challenge dataset show significant improvements over the state of the art without the need for self-ensembling, as well as improved data efficiency, reaching the performance of the challenge winner with just 25$\%$ of the training data.
\end{abstract}


%
\IEEEpeerreviewmaketitle

\section{Introduction}
\label{sec:intro}

Mapping the Earth with high resolution imagery is critical for a wide range of applications including environmental monitoring, urban mapping, disaster assessment, military surveillance, and many more. At the same time, instruments onboard of satellites face constraints such as payload sizes, downlink bandwidth, etc. that can limit the spatial resolution of the images they acquire, or the temporal availability of high-resolution (HR) products. Super-resolution (SR) techniques address the problem of estimating HR images from one or more low-resolution (LR) images. The availability of multiple images of the same scene is particularly useful since small geometric displacements allow the images to carry complementary information, that, when suitably combined by means of SR methods, can significantly increase the spatial resolution. In the context of remote sensing, there are several ways to obtain multiple images of the same scene, e.g., they can be acquired by a  spacecraft  during  multiple  orbits, or by multiple satellites imaging the same scene at different times, or may be obtained at the same time with different sensors. The most challenging scenario is the multitemporal one, as the content of the scene may change due to a variety of reasons, such as change in illumination, occlusions due to clouds, human activity, etc.

Significant progress has recently been made on multitemporal image SR, also spurred by the Proba-V challenge by the European Space Agency \cite{martens2019super, web:kelvins}. A curated dataset and a set of standard testing conditions allowed quick development of models tackling this difficult task and showed how deep learning models can effectively overcome difficulties such as unknown imaging model and temporal variation.

At the same time, we find that current models for multitemporal SR lack an important ingredient, which is the invariance to temporal ordering. Due to the unpredictability of change in the temporal series of LR images, no assumption can be made about patterns arising from a specific ordering of the images. In order to better understand this concept, let us compare it to a case where ordering matters, which is exploiting multiple spectral bands. In that case, each band has a specific physical meaning and ordering matters because there exist stable correlation patterns among bands. This is not the case when dealing with the temporal instead of spectral dimension, and any temporal permutation of the input LR images should always result in the same SR image. Therefore, we seek to build a model that is explicitly invariant to temporal permutation. Capturing this important prior allows to build a more robust and efficient model because it does not need to learn this property from the (possibly limited) data.

Moreover, in contrast with the LR images, for which established quality assessment criteria are typically available, the multitemporal SR process raises a problem related to the quality of the generated SR images, and, in particular, the degree of confidence to which a SR pixel has been reconstructed properly. If significant temporal variation is present, how can one trust the content of the SR scene to properly represent reality? We raise this issue for the first time in the context of deep multitemporal SR models and propose a technique that estimates an uncertainty value for each pixel in the SR image. We show that this uncertainty map correlates with temporal variations and with the true error signal. This uncertainty map can be made available to final users to judge the reliability of regions of the SR product.  

Our main novel contributions in this paper can be thus summarized as:
\begin{itemize}
    \item a new architecture for the multitemporal SR problem, called PIUnet (Permutation Invariance and Uncertainty network), which is invariant to temporal permutations and enables higher data efficiency, requiring smaller datasets for training;
    \item a method to estimate the aleatoric uncertainty of the SR image, consisting of an architectural design built in PIUnet and an ad-hoc training procedure;
    \item significant improvements over state-of-the-art on the Proba-V challenge dataset, in terms of both quality of the SR images and computational efficiency, since we do not require expensive temporal self-ensembles;
    \item a flexible model that can process an arbitrary number of input LR images in a stable manner and without ensembling or architecture redesigns.
\end{itemize}

\section{Related work}
\label{sec:relatedWork}

Image super-resolution has received great attention and a broad literature is available. However, the majority of works focus on single-image super-resolution (SISR) and a comparatively smaller number address the more challenging multi-image setting, and even fewer consider realistic LR degradations and multi-temporal change. 

SISR has been addressed by means of interpolation-based techniques, optimization-based methods \cite{ng2007total,6414620,6241428,Zhang2013SingleIS} and, more recently, learning-based methods relying on deep neural networks \cite{DnCnnZhang, liu2018non,10.1007/978-3-319-10593-2_13,kim2016deep,kim2015deep_rec,Shi2016RealTimeSI,Lim2017EnhancedDR,Zhang2018ResidualDN}. While SISR is interesting because of the limited amount of available information to solve the HR reconstruction problem, MISR offers a unique set of challenges and requires solutions that go beyond simple extensions of SISR works. 

The first work on MISR by Tsai and Huang \cite{tsaiHuang1984} used a frequency-domain technique to combine multiple downsampled images with subpixel displacements. However, frequency-domain algorithms do not allow to easily incorporate prior knowledge about HR images, and thus several spatial-domain MISR techniques were proposed over the years, including nonuniform interpolation \cite{1176931}, iterative back-projection (IBP) \cite{IRANI1991231}, projection onto convex sets (POCS) \cite{Stark:89,413332}, sparse coding \cite{Kato:2017:DSM:3066426.3066466, KATO201564}, and other regularized methods \cite{1331445,4060955,shen2009}. In particular, IBP \cite{IRANI1991231} enjoyed some success and works by improving the initial SR guess by back-projecting the difference between simulated LR images and actual LR images to the SR image, and iteratively attempting at inverting the forward imaging process. However, IBP is ultimately limited by the inability to deal with unknown or very difficult to model image degradation processes, as well as the difficulty in including image priors. Regularized methods generate the SR image by solving an optimization problem where a regularization cost can encode sophisticated image priors to improve performance. Among those, the bilateral total variation (BTV) method \cite{1331445} exploits a combination of the total variation regularizer and the bilateral filter to create a robust edge-preserving prior.

Model-based techniques such as the aforementioned ones are limited by the ability to accurately describe the forward imaging system and by the handcrafted prior used to capture the properties of real images. On the other hand, learning-based techniques directly use the data to overcome these modeling challenges and learn system models and image priors from observations. Deep neural networks are the tool of choice for such methods and recent years have shown progress in using deep learning for MISR. In particular, in the context of video SR \cite{7444187,DBLP:journals/corr/CaballeroLAATWS16}, convolutional neural networks (CNNs) have been developed to simultaneously perform motion compensation and SR frame generation. For instance, the dynamic upsampling filters (DUF) method proposed in \cite{Jo_2018_CVPR} estimates input-dependent interpolation filters for each pixel in the frames. Other applications of MISR can be found in burst photography \cite{bhat2021deep} where accurate registration and motion blur compensation play an important role. Recently, the Proba-V SR challenge \cite{martens2019super,web:kelvins}, issued by the European Space Agency, stimulated research for MISR approaches in the remote sensing context. The new dataset is particularly interesting for the development of new methods, as it allows to deal with realistic image degradation, registration problems, and robustness to temporal variation in the scenes, both among LR images and between LR and HR. The challenge winner DeepSUM \cite{molini2019deepsum} proposed a modular CNN composed of a SISR part, a module performing dynamic registration from the feature space and a fusion module based on 3D convolution. The architecture was further improved in DeepSUM++ \cite{molini2020deepsumpp} by using non-local operations in the form of graph-convolutional layers. The challenge runner-up HighRes-Net \cite{rarefin2020multi} proposed a recursive fusion strategy , also including a dynamic registration module. The current state-of-the-art is represented by the RAMS model \cite{salvetti2020multi}, which exploits feature attention at multiple stages.

 \begin{figure*}[ht]
\centering
\includegraphics[width=0.9\textwidth]{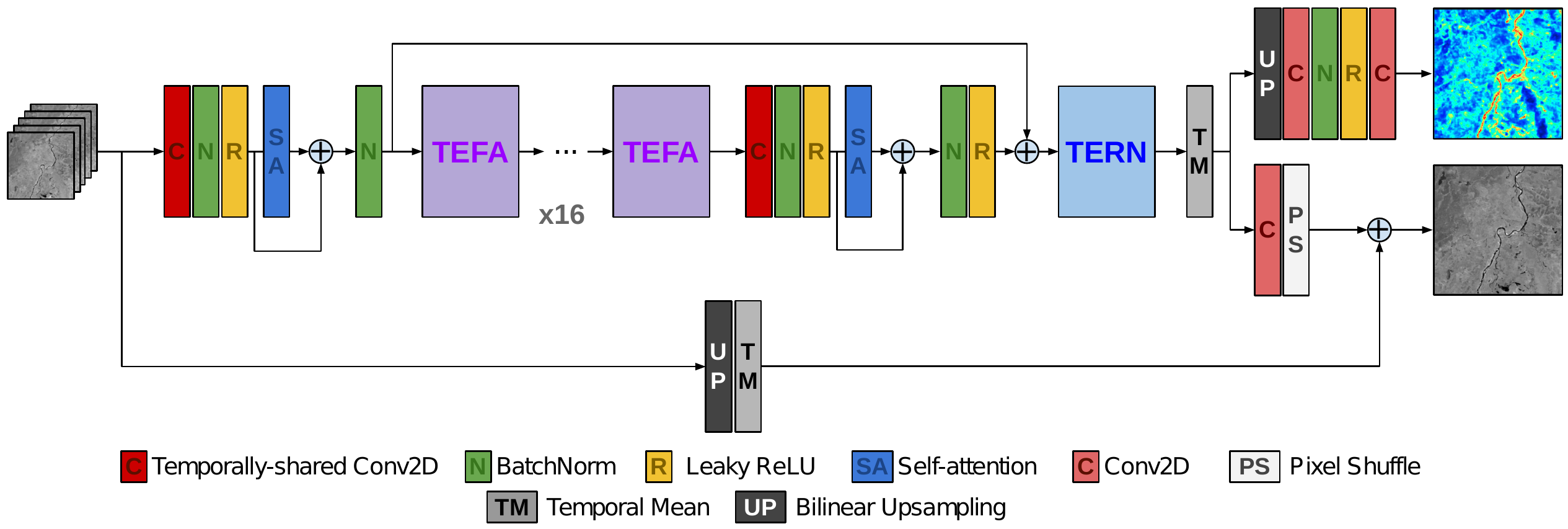}
\caption{PIUnet architecture. The model processes a stack of LR images and has two outputs, the top one being an uncertainty map and the bottom one the SR image.}
\label{fig:net}
\end{figure*}

\section{Proposed method}
\label{sec:method}

\begin{figure}[ht!]
\centering
\includegraphics[width=0.95\columnwidth]{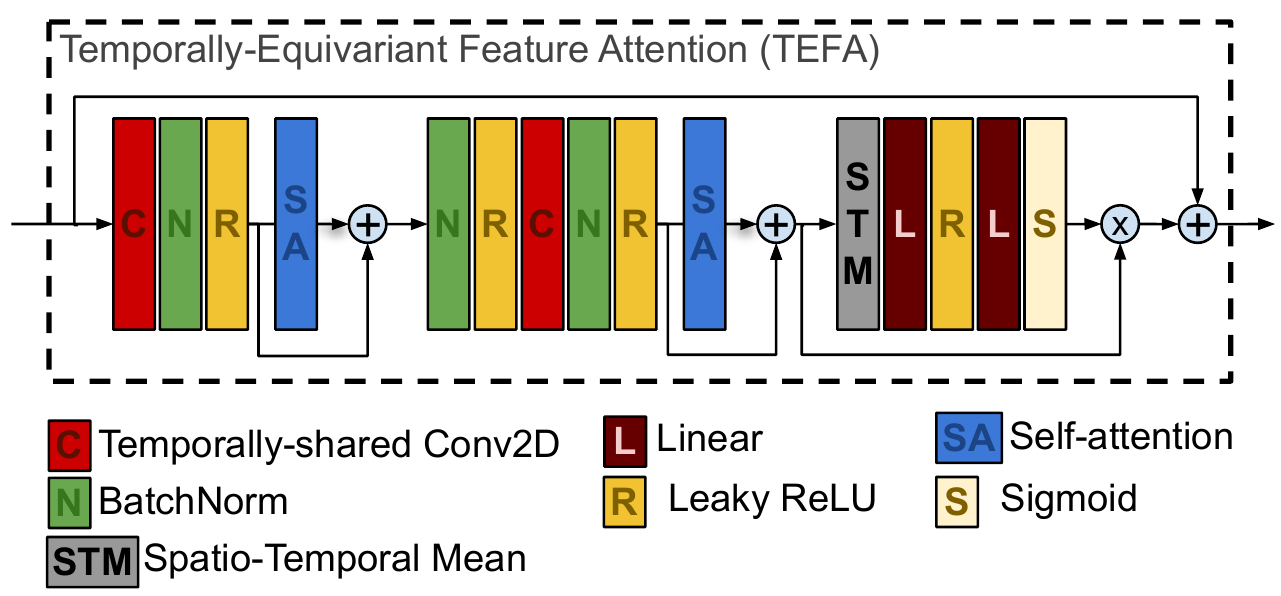}
\caption{Temporally-equivariant feature attention block.}
\label{fig:tefa}
\end{figure}

\begin{figure}[ht!]
\centering
\includegraphics[width=0.6\columnwidth]{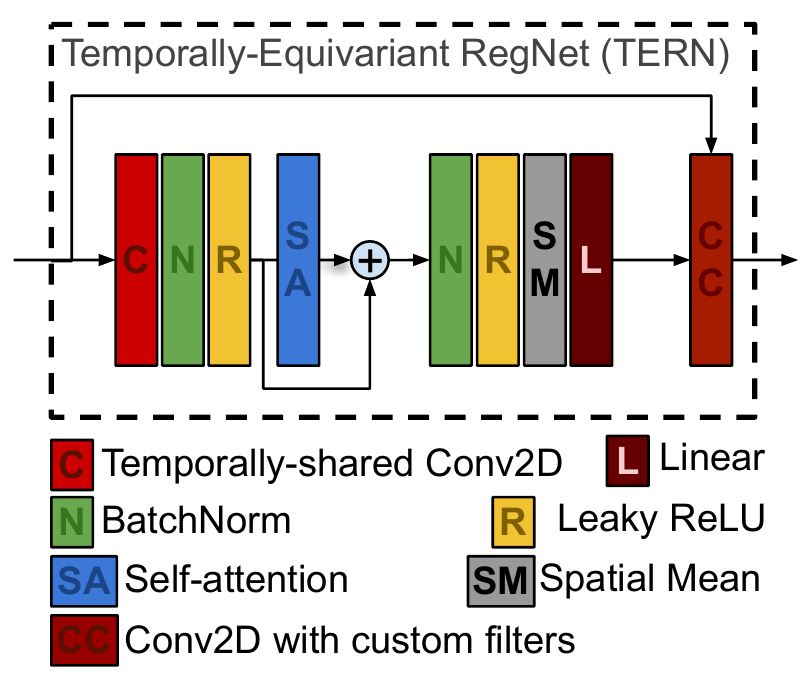}
\caption{Temporally-equivariant RegNet.}
\label{fig:tern}
\end{figure}

\subsection{Overview}

We propose to tackle multitemporal image super-resolution by means of a novel neural network design, called PIUnet, addressing Permutation Invariance and Uncertainty estimation. An overview is presented in Fig. \ref{fig:net}. The network input is a set with an arbitrary number of LR images. This is in contrast with other existing techniques which are designed for a fixed number of images. The LR images are assumed to be roughly registered with each other by means of preprocessing. The key features of PIUnet can be found in its invariance to permutations of the LR inputs and in the estimation of both the SR image and a corresponding pixel-by-pixel uncertainty value. This is achieved by means of two parallel heads which project the feature space built by the backbone to the two sets of information. Notice that the SR head uses pixel shuffling \cite{shi2016real} as upsampling technique and only estimates the residual from bilinear upsampling and averaging of the LR inputs, while the uncertainty head does not use this global skip connection. The next sections describe in detail how to achieve permutation invariance and how to estimate the uncertainty map.

\subsection{Invariance to temporal permutation}

In this section, we are going to discuss the main modules that allow the model to be invariant to permutations of the LR images in the temporal dimension. Before that, we need to introduce some terminology from group theory, and, in particular, the notions of equivariant and invariant functions.

\begin{definition}[Equivariance]
A function $f: X \rightarrow Y$ is said to be \textit{equivariant} to the actions $g$ of a group $\mathcal{G}$ if
\begin{align*}
    f(g \circ x) = g \circ f(x) \qquad \text{for all } x\in X, g \in \mathcal{G}.
\end{align*}
\end{definition} 

\begin{definition}[Invariance]
A function $f: X \rightarrow Y$ is said to be \textit{invariant} to the actions $g$ of a group $\mathcal{G}$ if
\begin{align*}
    f(g \circ x) = f(x) \qquad \text{for all } x\in X, g \in \mathcal{G}.
\end{align*}
\end{definition} 

In our case, we are dealing with the permutation group and its actions are all the possible temporal permutations of the input images. If we have an invariant function, the output will always be the same, no matter the permutation of the input, while for equivariant functions we will get an output that is exactly a permuted version of the output we would get without the input permutation. 

Invariance is a powerful property to exploit because it embeds robustness into the model, and allows to capture prior knowledge about the data properties, that would otherwise be very difficult to learn from the data. In our problem, multitemporal invariance is particularly desirable, as the SR image does not depend on any specific temporal ordering, and it would not make sense to assume that there exist ordered temporal patterns that always recur on every image set. Indeed, prior work on multitemporal super-resolution \cite{salvetti2020multi} attempted at capturing temporal invariance, but by means of data augmentation, where multiple permutations are fed to the model at training time, rather than by means of model operations. This augmentation approach is, at best, able to learn a weak invariance, as the model has to fully learn from the data that permutations correspond to the same output, rather than explicitly building this knowledge in the model operations. As an example, the fact that the training of the current state-of-the-art RAMS model \cite{salvetti2020multi} does not capture invariance is testified by the significant improvements obtained by test-time ensembling, where predictions corresponding to multiple temporal permutations are averaged.

Building a model out of invariant layers can be challenging because only a limited number of functions that are invariant to permutations exist, and since they are typically simple functions (e.g., the average) they may not be sufficiently expressive to build complex features. A simple technique to build an invariant model is, therefore, to use a sequence of equivariant operations, which are easier to design, followed by a global invariant function.
A trivial way to do that could be to independently process all the LR images with the same neural network and then combine the results with an invariant function, such as the mean. However, this is highly suboptimal because it does not exploit the correlation between images to build better feature spaces in the hidden layers. At the same time, using 2D convolution treating time as feature channels or 3D convolution would not provide equivariant operations, since the weights used by these layers assume a temporal ordering and any permutation would return a different result.

Therefore, we propose to use self-attention \cite{vaswani2017attention} in the temporal dimension as a permutation-equivariant operation that, at the same time, allows to effectively combine the information from the multiple time instants, exploiting their cross-correlations. We remark that, despite the name, self-attention is a significantly different operation from classic feature attention \cite{zhang2018image}. Self-attention projects its input feature vector to three different subspaces, using three learnable matrices, to generate the so-called key, query, value vectors and the cross-correlation matrix between key and query is used as transformation matrix to weigh the temporal components in the value vectors and generate the output. More formally, given the representation of a pixel $\mathbf{X} \in \mathbb{R}^{T \times F}$, characterized by $F$ features and $T$ temporal channels, the self-attention operation computes:
\begin{align*}
    \mathbf{Q}=\mathbf{X}\mathbf{W}_q,\quad \mathbf{K}=\mathbf{X}\mathbf{W}_k,\quad \mathbf{V}=\mathbf{X}\mathbf{W}_v\\
    \mathbf{Y}=  \mathrm{softmax}\left( \frac{\mathbf{Q}\mathbf{K}^\intercal}{\sqrt{T}}\right) \cdot \mathbf{V} = \mathbf{A} \mathbf{V}
\end{align*}
where the $\mathrm{softmax}$ function is applied row-wise, and $\mathbf{W}_q$, $\mathbf{W}_k$, $\mathbf{W}_v$ are learnable matrices.
Notice how this operation may look like a classic linear layer in the sense that the input is transformed by a matrix $\mathbf{A} \in \mathbb{R}^{T \times T}$, mixing the temporal channels, but the key difference is that $\mathbf{A}$ is computed as a function of the input itself, rather than being constituted of trainable values. It is easy to check that a permutation of the T temporal channels, results in a permutation of the columns of $\mathbf{A}$, ensuring the overall equivariance of the operation. This operation is performed for all pixels of all the images in our batch. We remark that self-attention has a quadratic complexity $O(T^2)$ in terms of computation and memory due to the $\mathbf{Q}\mathbf{K}^\intercal$ cross-correlations, but since we apply it to the temporal dimension, the value of $T$ is typically fairly small (e.g., in the Proba-V dataset, $T=9$), ensuring efficient implementations.

This operation provides us a building block to be used whenever we want to mix the temporal channels. Referring to Fig.\ref{fig:net}, we typically use 2D convolutions shared across the temporal channels to extract spatial features, and then use self-attention to temporally combine those features. Notice how sharing the 2D convolutions across the temporal dimension is crucial to maintaining equivariance to temporal permutation. Based on this idea, we also design a novel module to compute classic residual feature attention \cite{zhang2018image}, called Temporally-Equivariant Feature Attention (TEFA), shown in Fig. \ref{fig:tefa}, whose repetition serves as the backbone of our model. The classic residual feature attention \cite{zhang2018image} extracts a scalar value that is used to weigh each feature map. Our proposed TEFA module extends the idea to deal with the extra temporal dimension and computes attention scores to weigh the feature channels by extracting spatial and temporal features in an equivariant way, by means of the aforementioned shared 2D convolutions and temporal self-attention, and averaging them over space and time.

We also propose a temporally-equivariant extension of the RegNet module presented in DeepSUM \cite{molini2019deepsum}, called TERN and shown in Fig. \ref{fig:tern}. The goal of the original RegNet was to dynamically compute small $K \times K$ spatial kernels from the input features to be used as filters over the input itself. They served as adaptive filters that could implement registration filters, or, in general, spatial interpolators that could refine the registration of the multitemporal images, exploiting the powerful feature space of the network. For more details, we refer the reader to \cite{molini2019deepsum}. The original formulation was not perfectly equivariant as it relied on an explicit ordering where the first temporal image was taken as a reference and concatenated to each of the others to be processed in pairs as channels in a convolutional layer. In TERN, we overcome this limitation by exploiting self-attention to cross-correlate features over the temporal dimension and infer the values of the spatial kernels. Just like RegNet, TERN computes a different $K \times K$ spatial filter for each temporal image, while the filter is shared across multiple feature maps.

Finally, when we consider the proposed architecture from its input to the output of the TERN module, we can notice that it is equivariant to temporal permutations. In order to make the overall model invariant, we simply average the output of TERN along the temporal axis.

We also remark that the proposed model does not have any constraint on the temporal dimension, i.e., the same model could be used for any number of multitemporal images, which could be especially useful if, once deployed, fewer images were available. This is not the case for other existing methods, including DeepSUM \cite{molini2019deepsum} and RAMS \cite{salvetti2020multi}, which have architectures that have hard constraints on using exactly 9 multitemporal images, while only HighResNet \cite{rarefin2020multi} allows this flexibility thanks to their recursive fusion approach.

\subsection{Uncertainty estimation}

In this section, we propose a technique to assess a measure of confidence on the super-resolved image. This is important from the perspective of a final user of the SR product, whom we can inform about the degree of confidence a certain region of an image has to have been super-resolved properly.

We focus on characterizing the \textit{aleatoric} uncertainty \cite{kendall2017uncertainties} on the SR image. This kind of uncertainty is due to stochastic perturbations in the input data, typically noise, or, in this case, also temporal variations of the scene. Characterizing aleatoric uncertainty allows us to determine whether a portion of the image was poorly super-resolved due to variability in the input LR images. 

We use a heteroscedastic model for aleatoric uncertainty, essentially modeling each SR pixel as a random variable whose distribution can change on a pixel-by-pixel basis. The parameters of these distributions are then directly estimated by the neural network as its outputs. Training then uses the negative log likelihood (NLL) as loss function to be minimized. 
Most of the works on regression problems with neural networks \cite{kendall2018multi} model the samples as Normal random variables, thus the corresponding NLL can be seen as a generalization of the mean squared error (MSE) loss. However, it has been observed \cite{gallo2017loss} that, in many image restoration problems, the L1 loss typically outperforms the MSE loss, even if the evaluation metric is the tightly related PSNR. This has also been observed in the context of multitemporal super-resolution \cite{salvetti2020multi} and could be explained by the robustness of the L1 metric to outliers. We therefore seek an extension of the L1 loss to train our network, and this leads to modeling the pixel distribution as a Laplacian:
\begin{align*}
    p(x_i) = \frac{1}{2\beta_i}\exp\left( -\frac{\vert x_i-\mu_i \vert}{\beta_i} \right) \\
    \mathbb{E}\left[ x_i \right] = \mu_i, \quad \mathrm{Var}\left[ x_i \right] = 2\beta_i^2.
\end{align*}

The goal of the neural network is to output $\mu_i$, which will be the pixels in our SR image, and $\beta_i$ which is proportional to the standard deviation and will therefore be our aleatoric uncertainty. This is done with two parallel heads, as shown in Fig.1, splitting after the temporal averaging operation. In practice, $\delta = \log \beta$ is estimated for numerical stability, resulting in the following loss function:
\begin{align} \label{eq:nll}
    L &= - \frac{1}{NB} \sum_{b,i} \log p(x_i) \nonumber \\
    &= \frac{1}{NB} \sum_b \left[ \sum_i \left( \delta_i^{(b)} + e^{-\delta_i^{(b)}}\vert x_i^{\text{HR}(b)} - \mu_i^{(b)}  \vert \right) \right] 
\end{align}
for $i=1,\dots,N$ pixels and $b=1,\dots,B$ images.

It is often the case that the HR ground truth is not registered with the SR image. In that case, it is common practice \cite{rarefin2020multi,molini2019deepsum,salvetti2020multi} to use the minimum value of $L$ computed for all possible registration shifts as loss function.

We remark that estimating the uncertainty and using Eq. \eqref{eq:nll} as loss function actually serves a dual purpose. Not only it provides information on reliability of the SR image, but it also improves model performance with respect to the L1 loss. In fact, the contribution of the variance serves as a regularizer against excessively confident predictions, leading to higher quality solutions, as shown in the experiments in Sec. \ref{sec:nll_ablation}.

\begin{table*}[htb]
\centering
\caption{Quantitative performance - cPSNR (dB) and cSSIM}
\label{table:quantitative}
\begin{tabular}{lcccccccccc}
          & \multirow{2}{*}{Bicubic} & \multirow{2}{*}{IBP \cite{IRANI1991231}} & \multirow{2}{*}{BTV \cite{1331445}} & \multirow{2}{*}{DUF \cite{Jo_2018_CVPR}} & HighResNet & DeepSUM  & DeepSUM++ & \multirow{2}{*}{RAMS \cite{salvetti2020multi}} & RAMS \cite{salvetti2020multi} & \multirow{2}{*}{\textbf{PIUnet}} \\
          & & & & & \cite{rarefin2020multi} & (ens.) \cite{molini2019deepsum} & (ens.) \cite{molini2020deepsumpp} & & (ens.) &\\ \hline \hline
NIR cPSNR & 45.44 & 45.96 & 45.93 & 47.06 & 47.55 & 47.84 & 47.93 & 48.23 & 48.51 & \textbf{48.72}  \\
NIR cSSIM & 0.9771 & 0.9778 & 0.9794 & 0.9842 & 0.9855 & 0.9858 & 0.9862 & 0.9875 & 0.9880 & \textbf{0.9883}  \\
RED cPSNR & 47.34 & 48.21 & 48.12 & 49.36 & 49.75 & 50.00 & 50.08 & 50.17 & 50.44 & \textbf{50.62} \\
RED cSSIM & 0.9840 & 0.9865 & 0.9861 & 0.9842 & 0.9904 & 0.9908 & 0.9912 & 0.9913 & 0.9917 & \textbf{0.9921} \\ \hline
\end{tabular}
\end{table*}

\section{Experimental results and discussions}
\label{sec:results}

\begin{figure*}
    \centering
    \begin{subfigure}[b]{0.45\textwidth}
    \centering
    \includegraphics[width=0.8\textwidth]{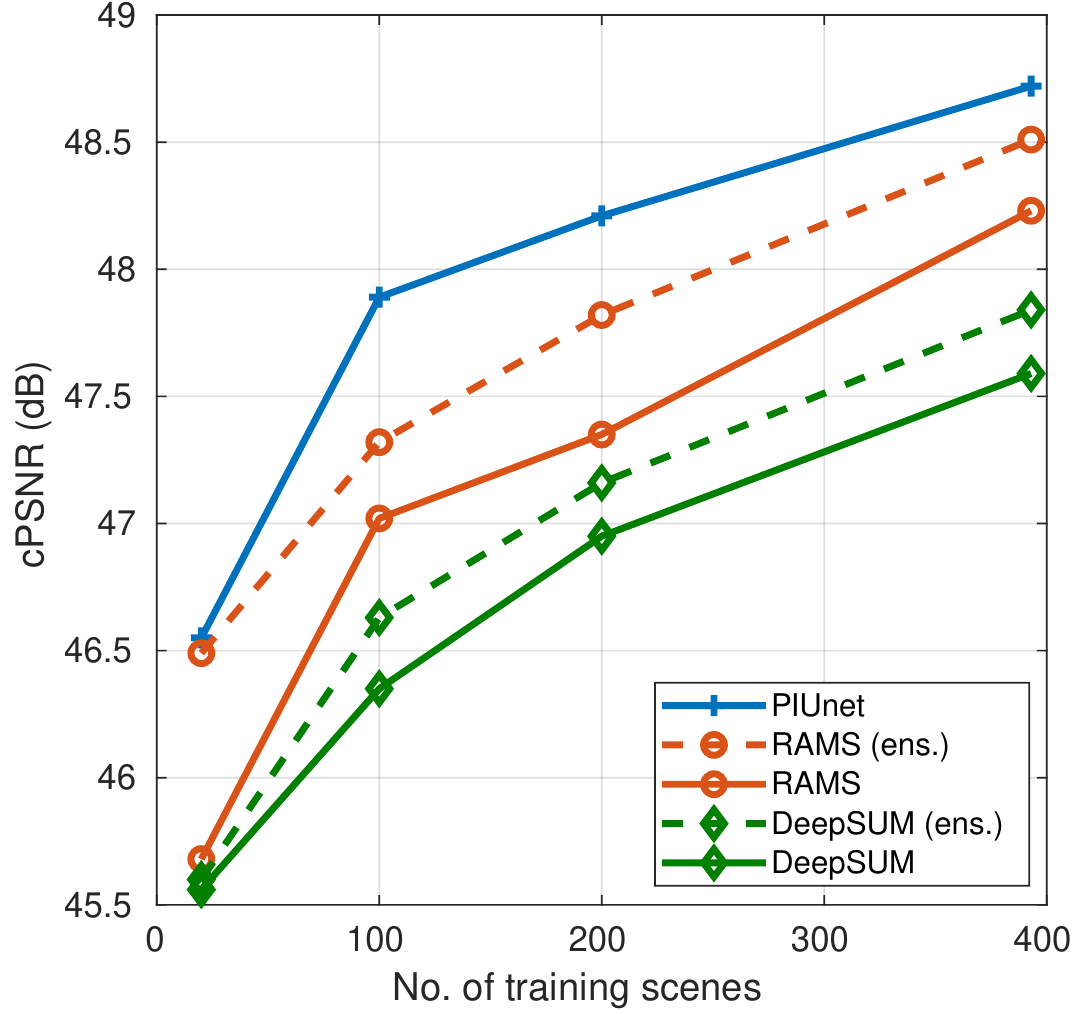}
    \caption{NIR}
    \end{subfigure}
    \begin{subfigure}[b]{0.45\textwidth}
    \centering
    \includegraphics[width=0.78\textwidth]{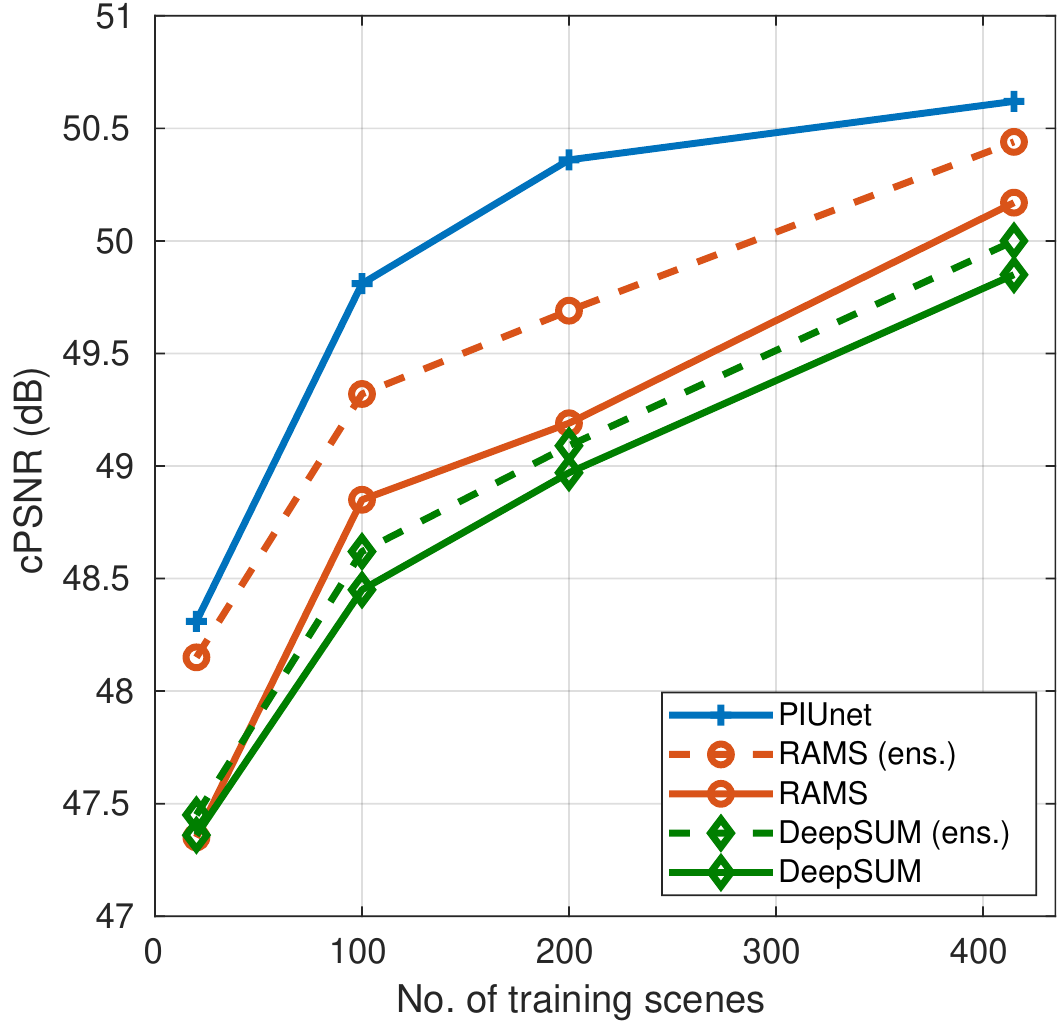}
    \caption{RED}
    \end{subfigure}
    \caption{Label efficiency. Performance on the validation set as function of the number of HR scenes available for training. Dashed lines represent temporal ensembles.}
    \label{fig:label_efficiency}
\end{figure*}

\begin{figure*}
    \centering
    \includegraphics[width=0.24\textwidth]{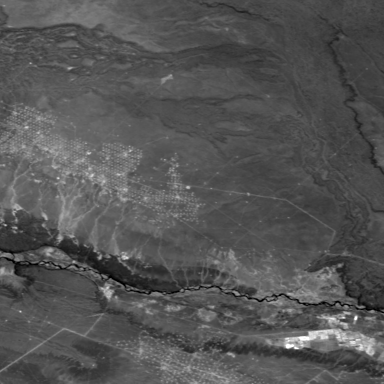}
    \includegraphics[width=0.24\textwidth]{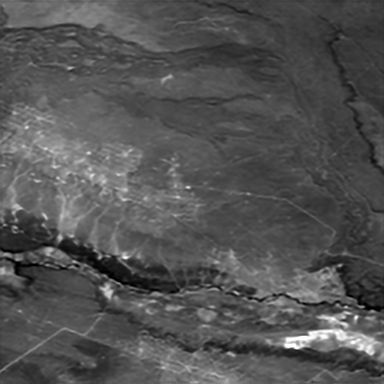}
    \includegraphics[width=0.24\textwidth]{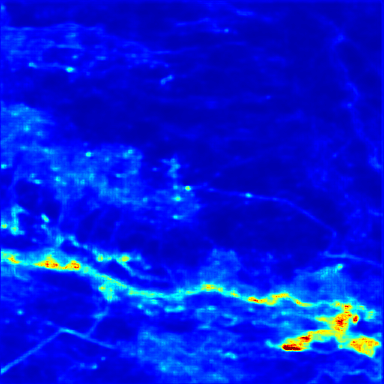}
    \includegraphics[width=0.24\textwidth]{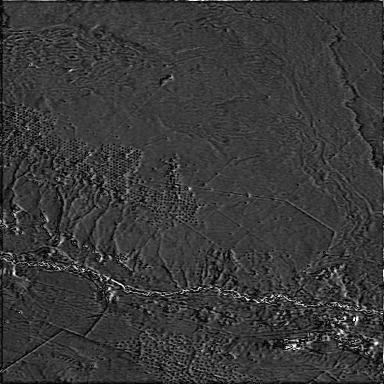}
    \caption{NIR validation \textit{imgset0792}. Left to right: HR, SR, SR uncertainty, ground truth absolute error.}
    \label{fig:output_example}
\end{figure*}

In this section, we discuss the experimental performance of the proposed method by focusing on the Proba-V dataset from the corresponding SR challenge. We present comparisons with the state-of-the-art models, including the most recent ones developed after the challenge. We also analyze the properties of the proposed method in terms of label efficiency, the impact of the Laplacian NLL loss, and the information provided by the uncertainty map. Code is available online: \url{https://github.com/diegovalsesia/piunet}.

\subsection{Experimental setting}
Our experiments employ the Proba-V super-resolution dataset \cite{martens2019super}, released by the European Space Agency in the context of a challenge \cite{web:kelvins}. The unique feature of this dataset is the availability of both LR and HR captured by the same satellite. In fact, Proba-V is able to acquire LR images at 300m resolution daily, but also HR images at 100m resolution with a longer revisit time of 5 days. The availability of real acquisitions for both LR and HR images makes for an interesting case study for SR techniques, since it avoids synthetic degradation of the HR data to generate the LR ones, which often results in simplistic degradations and distorts model performance. This is also well-suited for supervised learning techniques, which are able to fully learn an inverse to the complex and unknown degradation mapping. However, the higher revisit time for HR data means that only a limited amount of images could be available, once cloud cover and limits to temporal change are taken into account, so it is important to study efficient models that can perform well with few data. 
The dataset provides single-band Level 2A (radiometrically and geometrically corrected Top-of-Atmosphere  reflectance) images divided in a near-infrared (NIR) and visible (RED) categories. At least 9 LR images, acquired over the course of 30 days, are available for each HR scene. A total of 396 NIR scenes and 415 RED scenes are available for training, with additional 170 NIR and 176 RED scenes available for validation with known ground truth. The pixel size of the LR images is $128\times 128$, while the HR images are $384 \times 384$.

We preprocess the data by selecting only the LR images having cloud coverage lower than $15\%$ according to the provided clearance masks. We fix the number of used LR images per scene to 9 to match the standard setting used in other works. However, notice that the proposed model does not have any constraint on the size of the temporal dimension.  We also normalize the images by subtracting the average intensity over the training set and dividing by the standard deviation. LR images are registered to each other by means of cross-correlation, assuming a translational model. 
Training uses LR patches of size $32 \times 32$ extracted from all spatial locations in the available images and augmented with rotations, while testing directly processes the full pixel size, since the model is fully-convolutional. Separate models are trained for NIR and RED.

A feature size $F=42$ is used throughout the model, except for the linear layers in TEFA which form a bottleneck reducing to $F=5$ features before returning to 42. A total of 16 TEFA modules and 1 TERN module are used. The spatial kernels computed by TERN have size $5 \times 5$. The total number of trainable parameters is slightly under a million, and it is comparable with existing works. Training minimizes the NLL loss (Eq. \eqref{eq:nll}) for approximately 500 epochs, using the Adam optimizer \cite{kingma2014adam}. The NLL loss is made insensitive to misregistration between SR and HR and to absolute image brightness using the same approach as described in \cite{molini2019deepsum}:

\begin{align*}
    L &= \min_{u,v\in[0,6]} L_{u,v} \\
    L_{u,v} &= \frac{1}{NB} \sum_j \left[ \sum_i \left( \delta_i^{(j)} + e^{-\delta_i^{(j)}}\vert x_i^{\text{HR}(u,v,j)} - \hat{\mu}_i^{(j)}  \vert \right) \right] \\
    \hat{\mu}_i^{(j)} &= \mu_i^{(j)} + b^{(j)} \\
    b^{(j)} &= \frac{1}{\Vert m_i^{(j)} \Vert_1} \sum_j \left[ x_i^{\text{HR}(u,v,j)}m_i^{(j)} - \mu_i^{(j)}m_i^{(j)} \right]
\end{align*}
where $m_i^{(j)}$ is the clearance mask for image $j$, and $u,v$ indicate the amount of horizontal and vertical shift applied to the HR image.
The evaluation metric is the corrected PSNR, as used in earlier works, which accounts for shifts between the SR and HR images, and is insensitive to absolute brightness:

\begin{align*}
    \mathrm{cPSNR} &= \max_{u,v\in[0,6]} 10\log_{10} \frac{(2^{16}-1)^2}{\mathrm{MSE}_{u,v}}\\
    \mathrm{MSE}_{u,v} &= \frac{\Vert \mathbf{x}^{\text{HR}(u,v)}\odot \mathbf{m} - (\mathbf{x}^\text{SR}\odot\mathbf{m}+\mathbf{b}\odot\mathbf{m})) \Vert_2^2}{\Vert \mathbf{m} \Vert_1},
\end{align*}
with $\odot$ denoting elementwise product.

The learning rate is $10^{-4}$ and it is reduced to $2\cdot10^{-5}$ for the final epochs. A batch size equal to 24 was used. Training required approximately 2 days on a Titan RTX GPU.

\begin{figure*}[ht]
    \centering
    \begin{subfigure}[b]{0.39\textwidth}
    \includegraphics[width=0.935\textwidth]{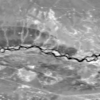}
    \end{subfigure}
    \begin{subfigure}[b]{0.6\textwidth}
    \includegraphics[width=0.3\textwidth]{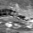}
    \includegraphics[width=0.3\textwidth]{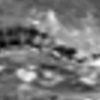}
    \includegraphics[width=0.3\textwidth]{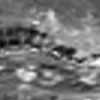}\\[2pt]
    \includegraphics[width=0.3\textwidth]{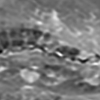}
    \includegraphics[width=0.3\textwidth]{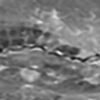}
    \includegraphics[width=0.3\textwidth]{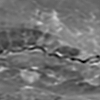}
    \end{subfigure}
    \caption{NIR validation \textit{imgset0792} detail. Left: HR image. Right from top left to bottom right: one among LR images, bicubic (47.71 dB/0.9874), IBP (48.46 dB/0.9892), DeepSUM (50.82 dB/0.9933), RAMS (51.24 dB/0.9939), PIUnet (51.81 dB/0.9946).}
    \label{fig:nir_qualitative}
\end{figure*}

\begin{figure*}[ht]
    \centering
    \begin{subfigure}[b]{0.39\textwidth}
    \includegraphics[width=0.935\textwidth]{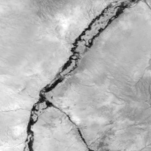}
    \end{subfigure}
    \begin{subfigure}[b]{0.6\textwidth}
    \includegraphics[width=0.3\textwidth]{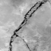}
    \includegraphics[width=0.3\textwidth]{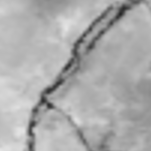}
    \includegraphics[width=0.3\textwidth]{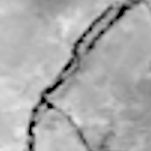}\\[2pt]
    \includegraphics[width=0.3\textwidth]{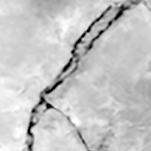}
    \includegraphics[width=0.3\textwidth]{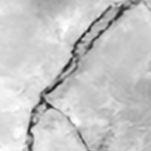}
    \includegraphics[width=0.3\textwidth]{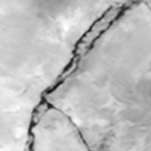}
    \end{subfigure}
    \caption{RED validation \textit{imgset0353} detail. Left: HR image. Right from top left to bottom right: one among LR images, bicubic (47.34 dB/0.9882), IBP (47.57 dB/0.9889), DeepSUM (48.78 dB/0.9913), RAMS (50.22 dB/0.9938), PIUnet (51.78 dB/0.9946).}
    \label{fig:red_qualitative}
\end{figure*}

\subsection{Comparison with state of the art}
In this section, we compare PIUnet with a number of approaches, representing the state of the art for the multitemporal SR task.  We include both model-based techniques and recent neural network methods to ensure a complete overview. The baseline (``Bicubic'') consists in bicubic upsampling, followed by pixel-domain registration and temporal averaging. For what concerns model-based approaches, we consider IBP \cite{IRANI1991231} and BTV \cite{1331445}. IBP takes  as  input  an  initial  guess corresponding to our Bicubic baseline and the spatial shifts related to the LR images using phase correlation algorithm. BTV also uses the same initial guess for the bilateral total variation minimization problem. Regarding neural network approaches, DUF \cite{Jo_2018_CVPR} is a technique from the video SR literature, using dynamic filters, a concept similar to what we use in TERN, but on a pixel-by-pixel basis; HighResNet \cite{rarefin2020multi} is the runner-up in the ESA challenge, and has a unique setting in which images in the LR set are recursively fused; DeepSUM \cite{molini2019deepsum} is the challenge winner and DeepSUM++ \cite{molini2020deepsumpp} an evolution with non-local operators based on graph convolution; RAMS \cite{salvetti2020multi} is the current state-of-the-art, heavily exploiting feature attention mechanisms.  

Table \ref{table:quantitative} show quantitative results in terms of cPSNR and SSIM for the various methods. Notice that the table reports ``(ens.)'' for results from published material obtained with self-ensembling techniques. In particular, DeepSUM (ens.) used ensembles over five subsets of 9 images, while RAMS (ens.) averaged the results of 20 temporal permutations of the input images. Highlighting the use of self-ensembling is important as it increases computational complexity, requiring more processing time and/or memory, with respect to a single-shot model. The results show that PIUnet outperforms the other methods, and, interestingly, even the ensembled version of RAMS.

We are also interested in testing the label efficiency of the methods that require training, i.e., what performance can be achieved if the training set only has a limited number of image with HR ground truth. The results are reported in Fig. \ref{fig:label_efficiency} which shows that PIUnet has significantly higher efficiency, by providing higher quality SR images even with constrained training sets. In particular, notice how just 100 scenes are required to reach approximately the same performance as DeepSUM (ens.).

Qualitative results are reported in Figs. \ref{fig:output_example}, \ref{fig:nir_qualitative}, \ref{fig:red_qualitative}. In particular, in Fig.\ref{fig:output_example}, notice how the output uncertainty map estimated by the PIUnet neural network correlates with the absolute error between the SR and HR images. Figs. \ref{fig:nir_qualitative}, \ref{fig:red_qualitative} show increased sharpness and higher fidelity with respect to the HR image achieved by the proposed method.

\begin{figure*}[ht]
    \centering
    \begin{subfigure}[b]{0.45\textwidth}
    \centering
    \includegraphics[width=0.8\textwidth]{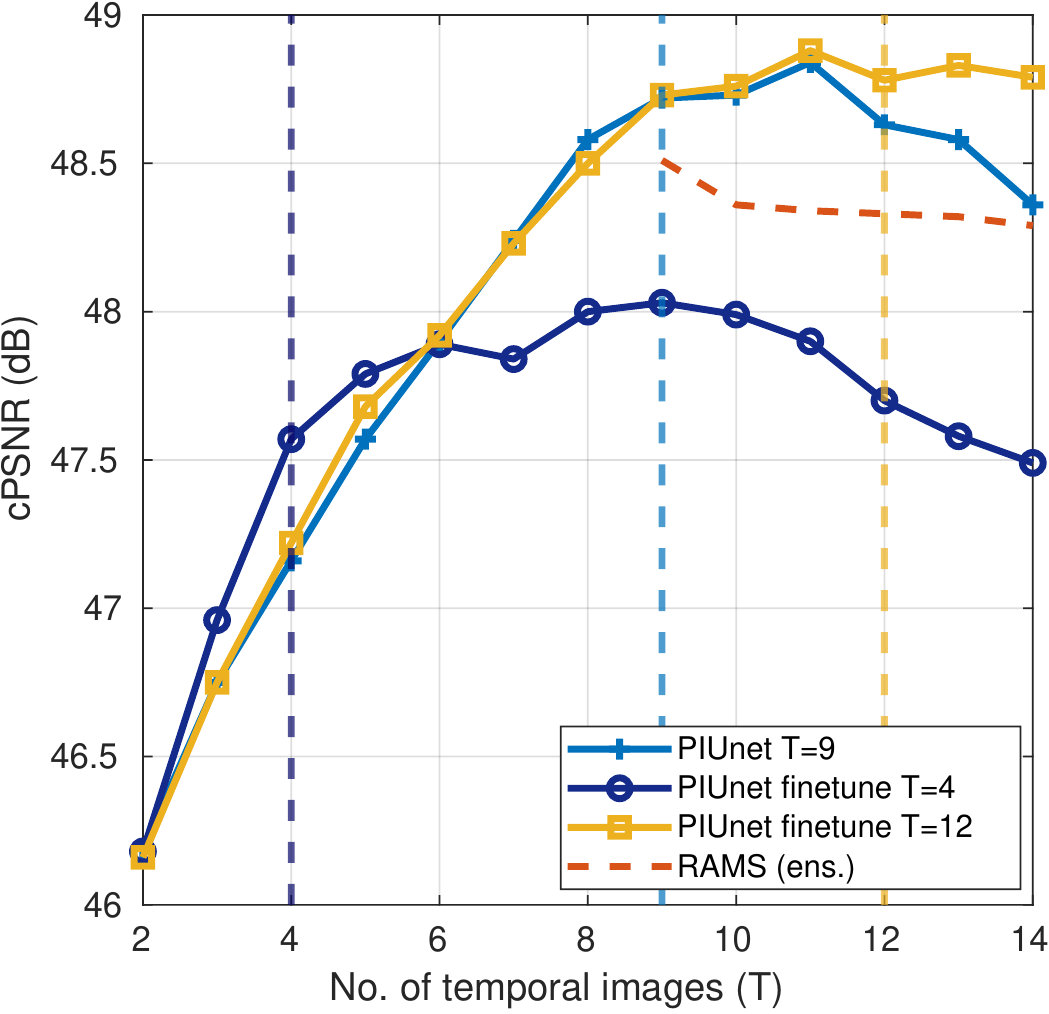}
    \caption{NIR}
    \end{subfigure}
    \begin{subfigure}[b]{0.45\textwidth}
    \centering
    \includegraphics[width=0.81\textwidth]{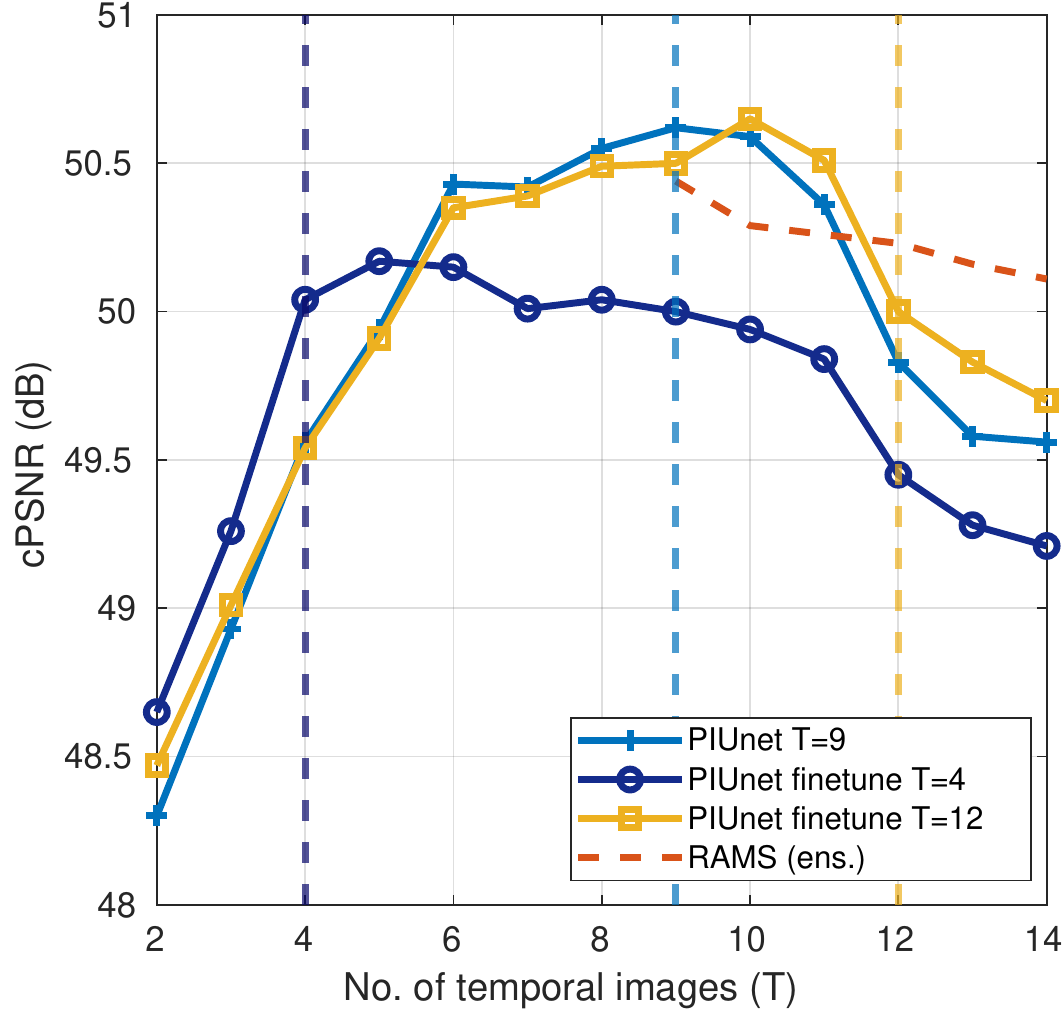}
    \caption{RED}
    \end{subfigure}
    \caption{Performance on the validation set as function of number of input LR images. Notice how the trained model gracefully handles any input size, but optimal performance may require finetuning.}
    \label{fig:n_images}
\end{figure*}

\subsection{Analysis and Ablation experiments}

\subsubsection{Number of input images}
We remark that PIUnet is capable of processing an arbitrary number of input images without resorting to self-ensembling techniques, which is a major limitation of state-of-the-art approaches like DeepSUM \cite{molini2019deepsum} and RAMS \cite{salvetti2020multi}. Moreover, those approaches cannot process a set of images which is smaller than nine without modifications to the architectures, due to the assumptions made on sizes in unpadded convolutional layers. It can be expected that PIUnet reaches optimal performance close to the number of images used during training. However, we expect a graceful behaviour as function of $T$. This is shown in Fig. \ref{fig:n_images}, where we can see that the best performance is achieved when the number of test images is close to (but not always at) the number of training images. The figure also shows that finetuning for a few iterations with the target number of images can improve performance, at the cost of degrading it for significantly different input sizes. Building a model with a broader optimality window so that finetuning is not required can be an interesting direction for future work.

\subsubsection{Impact of TERN module}
We evaluate the impact of TERN to the final performance of the model, and report it in Table \ref{table:tern}. This experiments replaces the TERN module with an extra TEFA module to keep the number of parameters approximately constant. We can notice that TERN provides significant performance improvements, especially on the NIR band.

\begin{table}[t]
    \centering
    \caption{Impact of TERN (cPSNR)}
    \label{table:tern}
\begin{tabular}{lcc}
    & TERN & TEFA  \\ \hline \hline
NIR & \textbf{48.72 dB} & 48.43 dB  \\
RED & \textbf{50.62 dB} & 50.55 dB \\
\hline
\end{tabular}
\end{table}

\subsubsection{Computational complexity}
Table \ref{table:complexity} reports some results on computational complexity of the proposed method with respect to state-of-the-art. We can see that PIUnet is significantly faster and with lower memory requirements than DeepSUM. The baseline version of RAMS is slightly faster and with comparable memory requirements. However, as shown in the previous section, we are able to outperform the temporally ensembled version of RAMS, which has a significant penalty in terms of complexity, as time or memory roughly scale linearly with the number of permutations in the ensemble. In particular, we report results for two ways of running the ensemble, i.e., serially, thus trading time for memory, or in parallel, trading memory for time. 
\begin{table}[t]
    \centering
    \caption{Computational complexity}
    \label{table:complexity}
\begin{tabular}{lcc}
                & runtime & memory  \\ \hline \hline
DeepSUM         & 484 ms & 5420 MB \\
RAMS            & 102 ms & 1250 MB \\
RAMS (ens.)     & 1642 ms / 1075 ms & 1250 MB / 5340 MB  \\
\textbf{PIUnet}           & 181 ms & 1280 MB \\
\hline
\end{tabular}
\end{table}

\begin{table}[]
    \centering
    \caption{Training loss comparison (cPSNR) }
    \label{table:nll}
\begin{tabular}{lcc}
    & L1 loss & NLL loss \\ \hline \hline
NIR & 48.41 dB  & \textbf{48.72} dB   \\
RED & 48.53 dB  & \textbf{48.62} dB   \\
\hline
\end{tabular}
\end{table}

\begin{figure*}[t]
    \centering
    \includegraphics[width=0.135\textwidth]{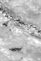}
    \includegraphics[width=0.135\textwidth]{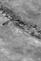}
    \includegraphics[width=0.135\textwidth]{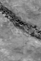}
    \includegraphics[width=0.135\textwidth]{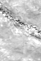}
    \includegraphics[width=0.135\textwidth]{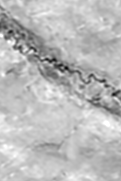}
    \includegraphics[width=0.135\textwidth]{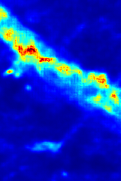}
    \includegraphics[width=0.135\textwidth]{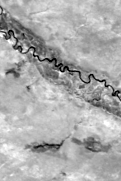}
    \caption{Temporal variation is captured by the SR uncertainty map. NIR validation \textit{imgset0975}. From left to right: four of the LR images, SR image, SR uncertainty map, HR image.}
    \label{fig:temporal_uncertainty}
\end{figure*}

\begin{figure*}[t]
    \centering
    \begin{subfigure}[b]{0.45\textwidth}
    \centering
    \includegraphics[width=0.8\textwidth]{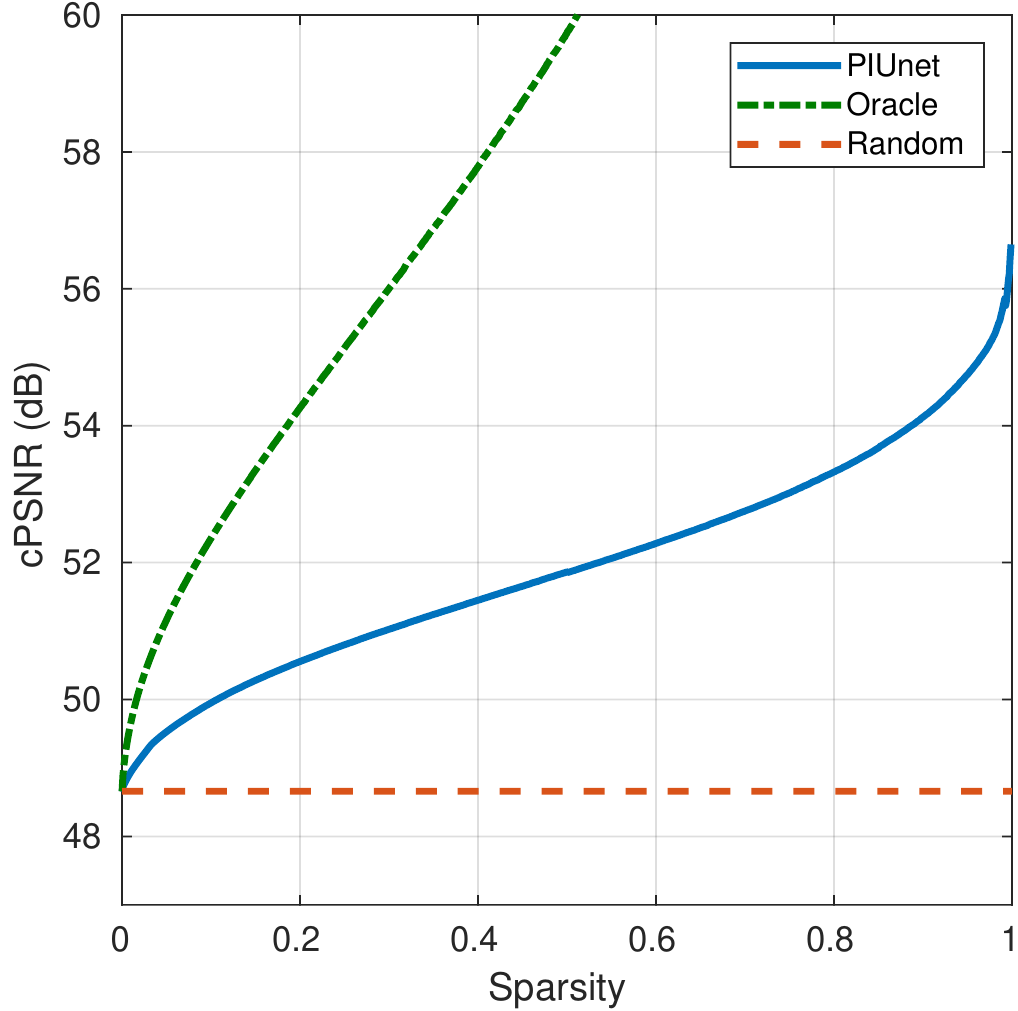}
    \caption{NIR}
    \end{subfigure}
    \begin{subfigure}[b]{0.45\textwidth}
    \centering
    \includegraphics[width=0.8\textwidth]{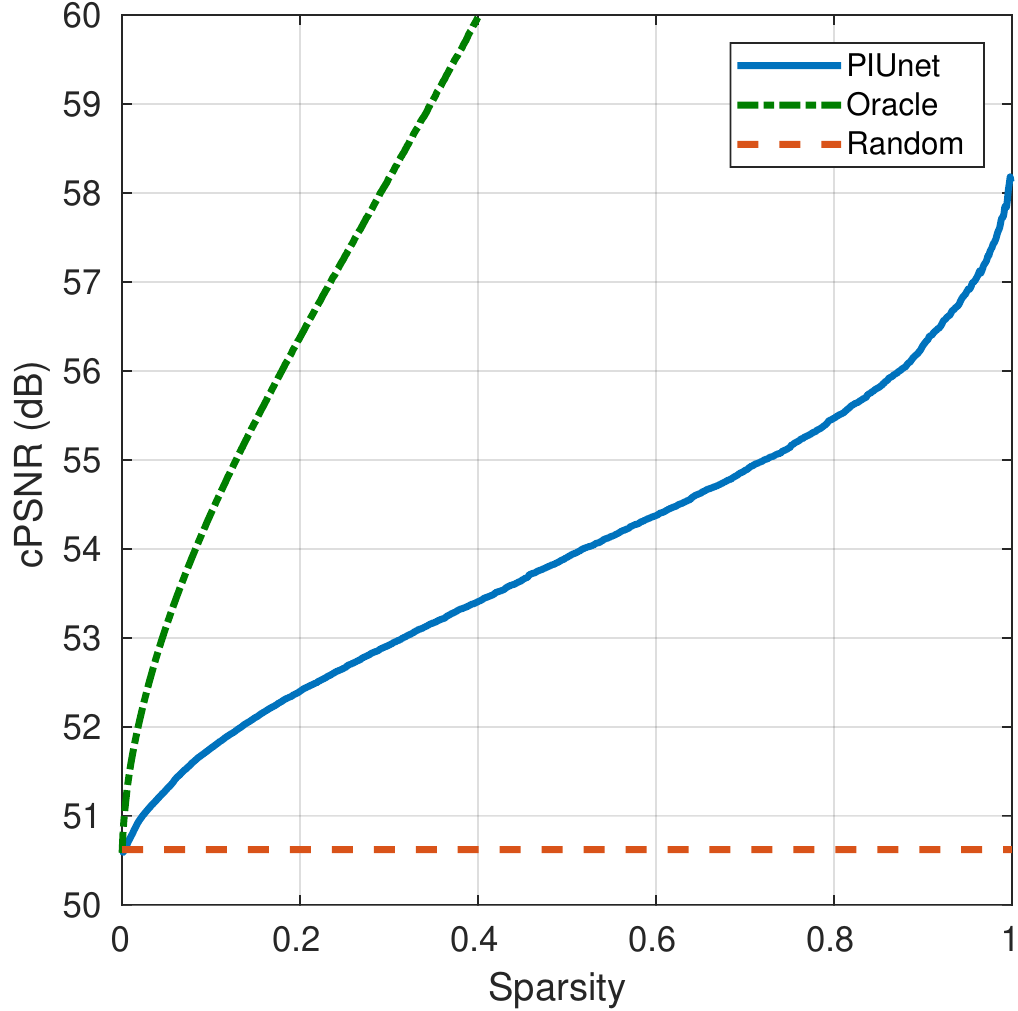}
    \caption{RED}
    \end{subfigure}
    \caption{Sparsification curves for the proposed method against random and oracle. The improvement over random shows that the output uncertainty map actually correlates with the error signal and provides valuable information on the local quality of the SR product.}
    \label{fig:sparsification}
\end{figure*}

\subsection{Uncertainty estimation}
\label{sec:nll_ablation}
In this experiment, we first show that using the NLL loss improves performance over the classic L1 loss. Table \ref{table:nll} shows the results obtained with the two losses for the same network architecture. We can see that the information on the variance of the super-resolved pixel helps convergence by regularizing high-confidence predictions and results in superior performance. This would already be a significant reason to use such loss, but an additional benefit is the availability of the uncertainty values on a pixel-by-pixel basis for the SR product. 

We can show how this uncertainty map can be used as guide to predict the unreliability of the SR image in areas with significant temporal variation. Fig. \ref{fig:temporal_uncertainty} shows an example of this concept. We focus on a crop of a scene displaying some temporal variation in the river bed and surrounding areas. Notice how the variance of the SR pixels is higher in the corresponding areas, highlighting to the user that the SR image might be less reliable due to the high content variability in the available images.

A quantitative way to determine whether uncertainty is correctly estimated is to check how it correlates with the true error map. This can be done by means of sparsification plots \cite{mac2012learning}. These plots are generated by first sorting all the pixels by decreasing uncertainty and then removing a progressively larger fraction of those with high uncertainty. If the uncertainty estimate correlates with the error signal, a quality metric measured on the remaining pixels should show improvements as more pixels are removed. Viceversa, if the uncertainty values were random, the curve would be a constant, as it does not provide informative ways of removing pixels to improve quality. Fig. \ref{fig:sparsification} shows the sparsification curve obtained for the proposed method, and compared with random selection and with an oracle that uses the true error signal to sort the pixels (for all variants, the brightness bias and shifts are computed before sparsification and kept fixed). The curves indeed confirm that the uncertainty map carries meaningful information. Future techniques addressing uncertainty estimation can compare sparsification curves to show improvements on the task.

\section{Conclusion}
\label{sec:conclusions}
We showed the importance of invariance to temporal permutation when building deep models for multitemporal image super-resolution. Building equivariant layers by means of self-attention and using a globally invariant operation close to the output allows to significantly improve performance over this challenging task and removes the computationally-expensive self-ensembling operation. We also showed how to estimate the uncertainty of the SR image, so that the final user can be guided in trusting the generated product, and that this further improves model performance.

\bibliographystyle{IEEEtran}
\bibliography{IEEEabrv,biblio}

\end{document}